\documentclass[a4paper,11pt]{article}
\usepackage{pos}
\usepackage{graphicx}
\usepackage{hyperref}
\usepackage{amsmath}
\usepackage{siunitx}

\begin{document}

\title{Novel Dark Matter Signatures}

\author[a]{A. Argiriou}
\author[b,c]{G. Cantatore}
\author[d]{S.A. Cetin}
\author[a, e]{E. Georgiopoulou}
\author[f]{D.H.H. Hoffmann}
\author[g]{S. Hofmann}
\author[h]{M. Karuza}
\author[i]{A. Kryemadhi}
\author[j]{M. Maroudas}
\author[k,l]{A. Mastronikolis}
\author[m]{E. L. Matteson}
\author[n]{K. \"{O}zbozduman}
\author[o,p]{Y. K. Semertzidis}
\author[a,q]{I. Tsagris}
\author[a,q]{M. Tsagri}
\author[r]{G. Tsiledakis}
\author[s]{E.L. Valachovic}
\author[t]{A. Zhitnitsky}
\author*[a]{K. Zioutas}

\affiliation[a]{Physics Department, University of Patras, 26504 Patras, Greece}
\affiliation[b]{Universit\`a di Trieste, 34127 Trieste, Italy}
\affiliation[c]{Istituto Nazionale di Fisica Nucleare (INFN), Sezione di Trieste, 34127 Trieste, Italy}
\affiliation[d]{Istinye University, Institute of Sciences, 34396 Sariyer, Istanbul, T\"{u}rkiye}
\affiliation[e]{Present Address: NTU Athens, Greece}
\affiliation[f]{Xi'An Jiaotong University, School of Physics, Xianning West Rd. 28, Xi'An, 710049, China}
\affiliation[g]{Independent Researcher, Fastlinger Strasse 17, 80999 M\"{u}nchen, Germany}
\affiliation[h]{Physics Department, University of Rijeka, 51000 Rijeka, Croatia}
\affiliation[i]{Physics Department, Messiah University, Mechanicsburg, PA 17055, USA}
\affiliation[j]{Institute for Experimental Physics, University of Hamburg, 22761 Hamburg, Germany}
\affiliation[k]{Department of Physics and Astronomy, University of Manchester, Manchester, UK}
\affiliation[l]{Present Address: Imperial College London, London, UK}
\affiliation[m]{Mayo Clinic College of Medicine and Science, Rochester, MN 55902, USA}
\affiliation[n]{Physics Department, Bogazici University, 34342 Bebek, Istanbul, T\"{u}rkiye}
\affiliation[o]{Center for Axion and Precision Physics Research, Institute for Basic Science, Daejeon 34051, Korea}
\affiliation[p]{Department of Physics, Korea Advanced Institute of Science and Technology, Daejeon 34141, Korea}
\affiliation[q]{Present Address: Geneva, Switzerland}
\affiliation[r]{IRFU/CEA-Saclay, CEDEX, 91191 Gif-sur-Yvette, France}
\affiliation[s]{Department of Epidemiology and Biostatistics, School of Public Health, University at Albany, State University of New York, New York, NY 12144, USA}
\affiliation[t]{Department of Physics and Astronomy, University of British Columbia, Vancouver, BC V6T 1Z1, Canada}

\emailAdd{marios.maroudas@cern.ch}

\abstract{Celestial observations often exhibit inexplicable planetary dependencies when the timing of an observable is projected onto planetary heliocentric positions. This is possible only for incident, non-relativistic streams. Notably, the celebrated dark matter (DM) in the Universe can form streams in our vicinity with speeds of about \SI[per-mode=symbol]{240}{\kilo\meter\per\second}. Since gravitational impact scales with $1/(\text{velocity})^2$, all solar system objects, including the Sun and the Moon, act as strong gravitational lenses, with their focal planes located within the solar system. Even the Moon can focus penetrating particles toward the Earth at speeds of up to approximately \SI[per-mode=symbol]{400}{\kilo\meter\per\second}, covering a large portion of the phase space of DM constituents. Consequently, the unexpected planetary dependencies of solar system observables may provide an alternative to \textit{Zwicky’s tension regarding the overestimated visible cosmic mass}. In this work, an overlooked but unexpected planetary dependency of any local observable serves as an analogue to Zwicky’s cosmic measurements, particularly if a similar mysterious behavior has been previously noted. Thus, a persistent, unexpected planetary dependency represents a new tension between observation and expectation. The primary argument supporting DM in line with Zwicky’s paradigm is this planetary dependency, which, on a local scale, constitutes the novel tension between observation and expectation. In particular, the recurrent planetary dependency of diverse observables mirrors Zwicky’s cosmic tension with the overestimated visible mass. No other approach accounts for so many otherwise striking and mysterious observations in physics and medicine.
}

\FullConference{2nd Training School and General Meeting of the COST Action COSMIC WISPers (CA21106) (COSMICWISPers2024)\\
10-14 June 2024 and 3-6 September 2024\\
Ljubljana (Slovenia) and Istanbul (Turkey)\\}

\maketitle

\section{Introduction}

From cosmological observations in tension with expectations, Zwicky (1933) \cite{Zwicky_Rotverschiebung} inferred that the total mass of the Coma galaxy cluster far exceeded its visible mass. By analyzing the cluster's galactic kinematics, he became the first to discover “dunkle Materie” (DM). This revealed an inconsistency in known physics with profound implications that persist today. We now understand that our Universe is dominated by DM, an as-yet mysterious substance. As its name suggests, DM is dark, meaning it does not emit or reflect light, carries no electric charge, and interacts extremely weakly with normal matter, making it undetectable with conventional instruments \cite{Liu_Photonic}.

However, this definition may be misleading, as several local counter-observations (some presented below) could be attributable to DM. Like Zwicky’s discovery, these observations challenge known physics, despite the commonly accepted DM framework. Notably, diverse local findings \cite{Zioutas_invisible, Zioutas_melanoma, Zioutas_response, Zioutas_27_melanoma, Zioutas_births} prompt the question: can Zwicky’s reasoning be applied locally? To explore this, we identify multiple striking tensions that restore his logic on a smaller scale. A key example is the unexpectedly observed planetary dependencies. This is because the only known remote planetary tidal force is extremely feeble, incapable of causing visible impacts on the dynamical behavior of solar system bodies, including the Sun and the Earth. Yet, observations suggest the presence of an unexplained planetary force. Interestingly, for DM streams or clusters \cite{Vogelsberger_streams, Kryemadhi_gravitational, Tkatsev_Bose_star, Kolb_miniclusters} with a velocity distribution around \SI[per-mode=symbol]{240}{\kilo\meter\per\second}, most solar system bodies act as efficient gravitational lenses \cite{Hoffmann_Gravitational, Patla_flux, Sofue_gravitational, Prezeau_hairs}, with focal lengths within the solar system, including the Moon focusing DM particles toward the Earth. For instance, during the alignment of a DM stream with an intervening solar system body, the enhanced DM flux downstream may cause a spatiotemporal increase in the interaction rate, far exceeding the mean value usually regarded as background—or it may even go unnoticed.

\begin{figure}[!htb]
    \centering
    \includegraphics[width=0.55\linewidth]{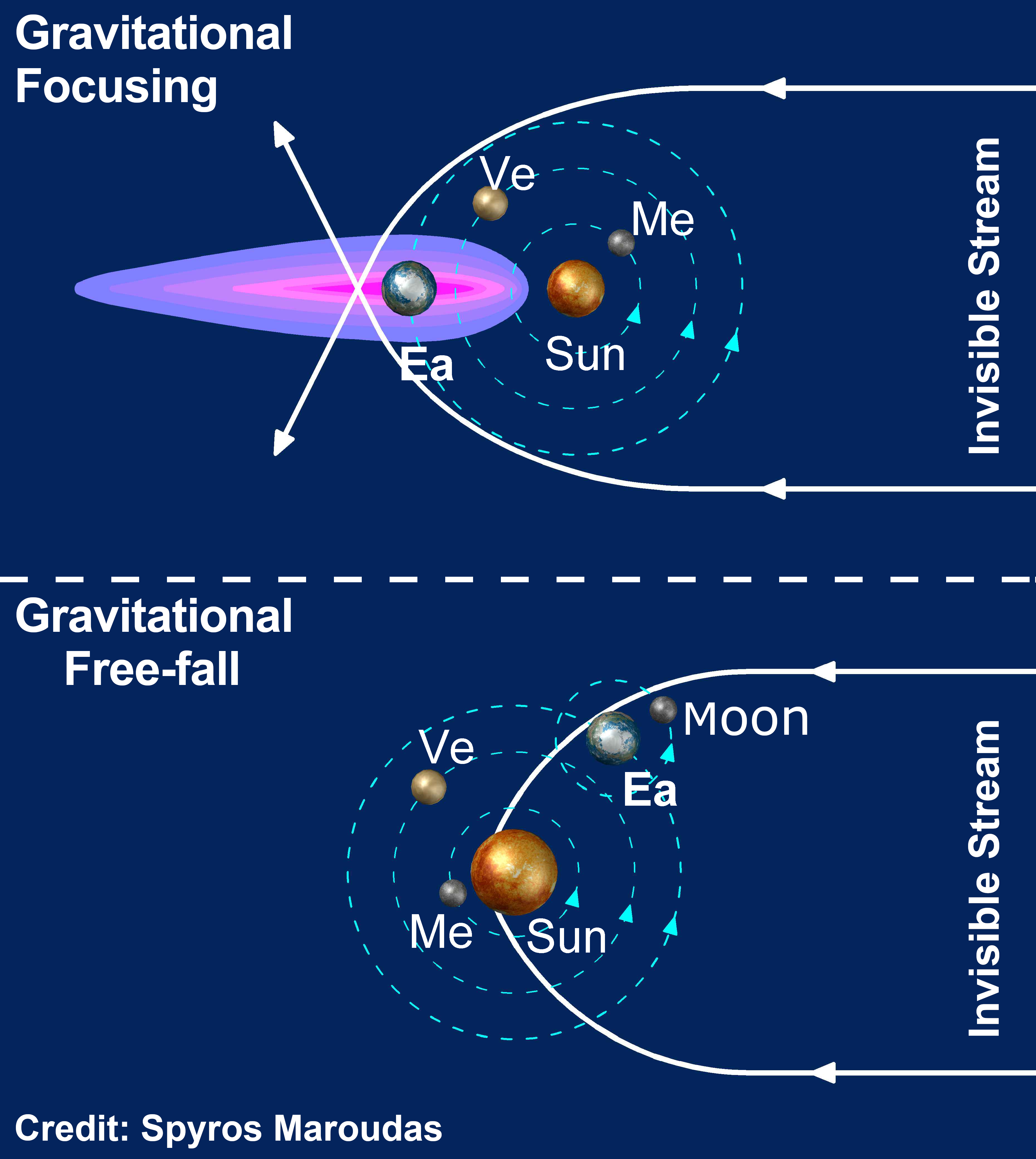}
    \caption{Graphic representation of gravitational focusing effects of an invisible DM stream by the solar system. Top: gravitational focusing by the inner solar system. In this configuration, the galactic center is on the right side, in the opposite direction of the incident invisible stream. Bottom: the free-fall effect of incident low-speed streams also may dominate planetary gravitational focusing towards the Sun since the flux enhancement increases with $(v_\text{escape}/ v_\text{incident})^2$ where $v_\text{escape}$ is the escape velocity from the Sun and $v_\text{incident}$ the initial particle velocity far away from the Sun. The flux towards Earth can also be gravitationally modulated by intervening planets. The Moon focuses particles towards the Earth with an incident velocity near the Moon up to \SI[per-mode=symbol]{400}{\kilo\meter\per\second} \cite{Zioutas_invisible, Vogelsberger_streams, Kryemadhi_gravitational, Tkatsev_Bose_star, Kolb_miniclusters}.}
    \label{fig:grav_focusing}
\end{figure}

The driving idea behind this work is as follows: planetary gravitational effects (including those of the Sun and the Moon) on non-relativistic “invisible massive particles” can focus these particles on solar and planetary atmospheres (see Fig.~\ref{fig:grav_focusing} and references \cite{Hoffmann_Gravitational, Patla_flux}). Throughout this work, we occasionally refer to “invisible matter” to broaden the DM horizon beyond celebrated candidates like axions and WIMPs. If DM constituents interact significantly with normal matter or radiation, they could already have observable impacts on outer atmospheres, such as the Sun’s or planetary ionospheres and stratospheres. Despite their differences, these atmospheres exhibit striking anomalies: the solar corona paradox, long-known ionospheric dynamical anomalies, and the recurring upper stratospheric temperature excursions in early January \cite{Bertolucci_sun, Zioutas_stratospheric}. Atmospheric layers could screen possible DM signatures, modifying observations below—potentially even affecting underground DM detection experiments. Furthermore, strong planetary correlations have been observed in inner Earth dynamics, such as the unexplained planetary dependence of earthquakes (EQs) \cite{Zioutas_atmospheric}, when their timing is projected onto planetary orbital positions (see Figs.~6 and 7 in \cite{Zioutas_atmospheric}).

A viable scenario considered in this work involves planetary alignment with an incident invisible stream. The existence of DM streams has been proposed independently of this work, based on cosmological considerations \cite{Vogelsberger_streams}. A potential signal should repeat if the DM stream persists for periods much longer than planetary orbital periodicities, which is reasonable to expect. Notably, the orbital periodicities of a single planet or synods of two or more planets often result in spatiotemporally peaked signal enhancements, rather than a time-averaged washed-out effect. For example, the triple synod of Jupiter, Earth, and Venus remarkably coincides with the 11-year solar cycle, which remains a mystery within known physics. This coincidence is most likely not random and supports the proposed streaming DM scenario, as discussed in \cite{Zioutas_invisible} and corroborated by several follow-up observations, including in vivo studies \cite{Zioutas_melanoma, Zioutas_response, Zioutas_27_melanoma, Zioutas_births}.

If similar planetary correlations appear in exo-solar planetary systems \cite{Perryman_gaia}, it would provide strong independent confirmation of streaming DM. This would imply that orbiting exo-planetary gravitational lenses also act as gravitational lenses focusing DM constituents onto host stars or other celestial bodies regardless of the nature of DM. Due to the dependence of the lensing focal length on $1/(\text{velocity})^2$, even the Moon can focus DM particles toward the Earth at velocities up to approximately \SI[per-mode=symbol]{400}{\kilo\meter\per\second}, covering a significant fraction of the DM velocity phase space \cite{Zioutas_invisible, Kryemadhi_gravitational, Sofue_gravitational, Prezeau_hairs}. The same logic extends to exo-solar systems.

This work highlights several solar and terrestrial conventionally unexpected observations, including a long series of medical data on diagnosed melanomas (a type of skin cancer) \cite{Zioutas_melanoma, Zioutas_response, Zioutas_27_melanoma}, as well as similar multiple planetary signatures in non-malignant living matter \cite{Zioutas_births}, i.e., in vivo measurements. One or more planetary correlations in any observable could serve as a novel signature of the dark sector, given the absence of any known remote planetary force beyond the extremely feeble and smooth tidal force \cite{Javaraiah_longterm}. For the streaming DM scenario, the gravitational focusing plane of an invisible stream depends on $1/(\text{velocity})^2$ \cite{Hoffmann_Gravitational}, greatly favoring flux enhancements within the solar system for the non-relativistic speeds typical of DM constituents. The usually accepted global mean value of the DM density ($\sim\SI[per-mode=symbol]{0.45}{\giga\electronvolt\per\centi\meter\cubed}$) can experience spatiotemporal excursions by orders of magnitude.

Occasional planetary gravitational focusing effects may lead to enormous DM flux enhancements within the solar system. This is feasible only if invisible matter consists, at least partly, of DM streams with velocities typical of the dark sector (\SI[per-mode=symbol]{240}{\kilo\meter\per\second}). Notably, following \cite{Vogelsberger_streams}, as many as $10^{14}$ “fine-grained” cosmic streams may exist in our galaxy, exposing the solar system to approximately $10^6$ or more streams \cite{Kryemadhi_gravitational}. Streaming “invisible matter” emerges as the only viable explanation (see \cite{Zioutas_invisible, Kryemadhi_gravitational}) for occasionally observed diverse anomalies in our vicinity, such as the enigmatic 11-year solar cycle. The cosmologically derived streaming DM scenario \cite{Vogelsberger_streams} aligns with phenomenological proposals (see subsect.~\ref{11yrs_cycle}), despite their distinct origins. Both approaches converge on the existence of streaming DM, further reinforcing its plausibility.

As Frank Wilczek (Nobel 2004) emphasized during a seminar at CERN: “Focus on anomalies and mysteries.” By projecting their timestamps onto planetary orbital positions, multifaceted solar mysteries align well with Wilczek’s recommendation.

\section{Local DM Signature \'{a} la Zwicky}

The discovery of DM stemmed from a fundamental conflict with known physics—specifically, cosmic gravity. In 1933, Zwicky observed the Coma galaxy cluster and found that its visible matter was insufficient to account for the observed gravitational effects, revealing a striking discrepancy \cite{Zwicky_Rotverschiebung}. This work is motivated by a key question: can Zwicky’s logic also be applied within our solar system? If persistent and significant local observations strongly contradict expectations, then, following Zwicky’s reasoning, an external influence must be responsible. Could DM be the culprit? The answer seems inevitable—what else? The various phenomena discussed in this work cannot be attributed to any nonexistent remote planetary force, making streaming DM the most plausible unifying explanation.

The unexpected planetary dependencies observed in multiple anomalies introduce a novel paradigm \textit{à la Zwicky}. Notably, planetary tidal forces acting on the Sun are approximately 12 orders of magnitude too weak \cite{Javaraiah_longterm} to exert any significant remote influence. Yet, observational data consistently reveal planetary correlations, necessitating an alternative explanation—one that aligns with DM interactions rather than conventional gravitational forces, which should exhibit a smooth temporal behavior over long periods of time.

\section{Local Observations}

Numerous solar and terrestrial observations exhibit unexpected planetary correlations, each with a statistical significance exceeding $5\sigma$. Most of these studies reveal multiple planetary dependencies, further reinforcing their validity. Many of these phenomena have long been considered anomalies or unresolved mysteries within known physics, including the unnaturally hot solar corona, solar flares, ionospheric plasma density variations, annual temperature fluctuations in the upper stratosphere, and biomedical findings related to melanoma and non-malignant living matter in vivo \cite{Zioutas_melanoma, Zioutas_response, Zioutas_27_melanoma, Zioutas_births}.  

These conventionally unexpected planetary correlations serve as a common footprint supporting the reasoning presented in this work, analogous to Zwicky’s discovery of the cosmic mass discrepancy. Here, planetary dependencies take the place of Zwicky’s tension between observed and expected visible matter on a cosmic scale. In the following sections, we discuss published results presented in detail mainly in the PhD thesis of M. Maroudas \cite{Maroudas_phd}, providing corresponding references with online access where possible to facilitate verification while avoiding copyright concerns.

In the present work, we analyze planetary dependencies within the Fourier periodogram framework. In reality, these dependencies are identified by projecting the timestamps of an observable onto heliocentric longitudes. However, the use of the Fourier spectra facilitates the comparison of different observables, allowing us to assess whether correlations between two or more observables exist or not.

\subsection{The 11-Year Solar Cycle}
\label{11yrs_cycle}

The 11-year solar cycle remains one of the most profound unresolved mysteries in known physics \cite{Zioutas_11years}, yet it is closely tied to the dynamical behavior of the solar system. The significance of this cycle was a key motivation for the research presented in this work. Notably, this periodicity coincides with the triple synod of Jupiter, Earth, and Venus. Given the established planetary dependencies within the solar system, the recurrence of this triple synod in sync with the 11-year cycle is unlikely to be coincidental.

\begin{figure}[!htb]
    \centering
    \includegraphics[width=\linewidth]{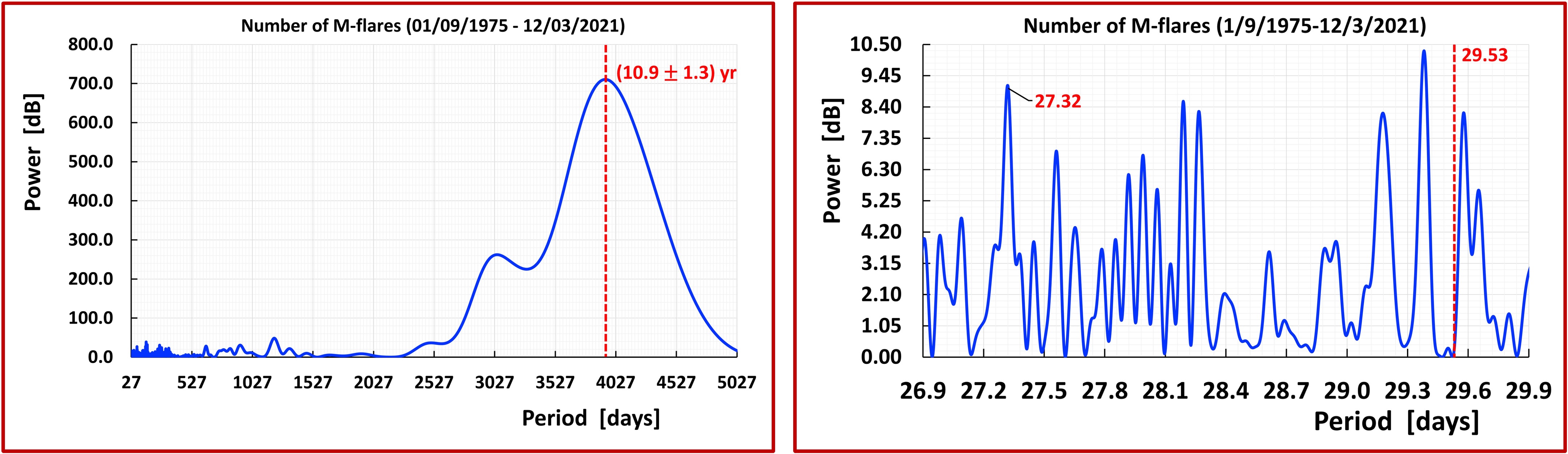}
    \caption{Fourier periodograms of M-class solar flares from 01/09/1975 to 12/03/2021. \textbf{Left}: Periodicity analysis over a range of 27 to 5027 days, clearly revealing the presence of the 11-year solar cycle \cite{Bertolucci_sun}. \textbf{Right}: Zoomed-in view of the periodicity range from 27 to 30 days, highlighting peaks at 27.32 days (lunar sidereal period) and 29.53 days (lunar synodic period). The prominence of the sidereal peak suggests that solar flares may have an origin beyond the solar system \cite{Bertolucci_sun}.}
    \label{fig:m_flares}
\end{figure}

Fig.~\ref{fig:m_flares} (left) presents the Fourier spectrum of M-class solar flares recorded from 1975 to 2021, clearly revealing the presence of the 11-year cycle. This cycle also governs the occurrence of sunspots, which exhibit intense magnetic fields in the kilogauss range. Solar magnetism is central to the Sun’s activity, and the intensity of X-ray emissions from sunspots follows a quadratic dependence on the local magnetic field strength \cite{zioutas_indirect}.

\subsection{Solar Flares}

The first solar flare was observed by Carrington in 1859 \cite{Carrington_description}. Since then, the underlying mechanisms that power and trigger solar flares remain unexplained by current physics. Could their unpredictability be rooted in more fundamental causes? As with many solar phenomena, flares follow the enigmatic 11-year solar cycle (Fig.~\ref{fig:m_flares}, left), which intriguingly coincides with the triple synod of Jupiter, Earth, and Venus. This alignment is unlikely to be coincidental, suggesting a planetary influence on solar activity, with sunspot and flare frequencies serving as key proxies.

The Fourier periodogram in Fig.~\ref{fig:m_flares} (right) highlights a notable peak at 27.32 days, corresponding to the lunar sidereal period, which is tied to remote stars. This additional periodicity strengthens the involvement of an external (exo-solar) factor beyond the well-known 11-year cycle which coincides with ac combined triple planetary dependency. The concurrence of both the 11-year solar cycle and the 27.32-day lunar cycle with specific planetary configurations—such as the 11-year Jupiter-Earth-Venus synod and the 237-day Jupiter-Venus synod \cite{Zioutas_births}—strongly suggests a planetary connection to solar flare activity. 

\subsection{The Solar Corona}

The solar corona heating problem remains a significant unresolved issue in conventional astrophysics. Specifically, the Sun's outer atmosphere is roughly 100–1000 times hotter than its surface. Current explanations point to unknown physical processes that govern this heating, with some suggesting that solar X-ray emission above active regions is influenced by magnetic fields. However, these explanations fail to fully account for the extreme temperatures observed. Could an unseen corona exist, or might the heating result from secondary effects of unknown processes? The question of “unobserved physical processes governing the heating...” remains open.

Interestingly, solar radiation above $\sim \SI{25}{\eV}$ should not exist according to blackbody radiation expectations, as the Sun should have thermally relaxed over its $\sim$4.5 Gyr lifetime \cite{Judge_intermittency}. By comparison, the entire Universe reached thermal equilibrium within a few hundred thousand years. Notably, the unexpected excess solar radiation, primarily in the UV and EUV range ($\sim$25 eV to $\sim$1–2 keV), displays a planetary dependency \cite{Bertolucci_sun}, adding another layer of mystery.

\begin{figure}[!htb]
    \centering
    \includegraphics[width=0.65\linewidth]{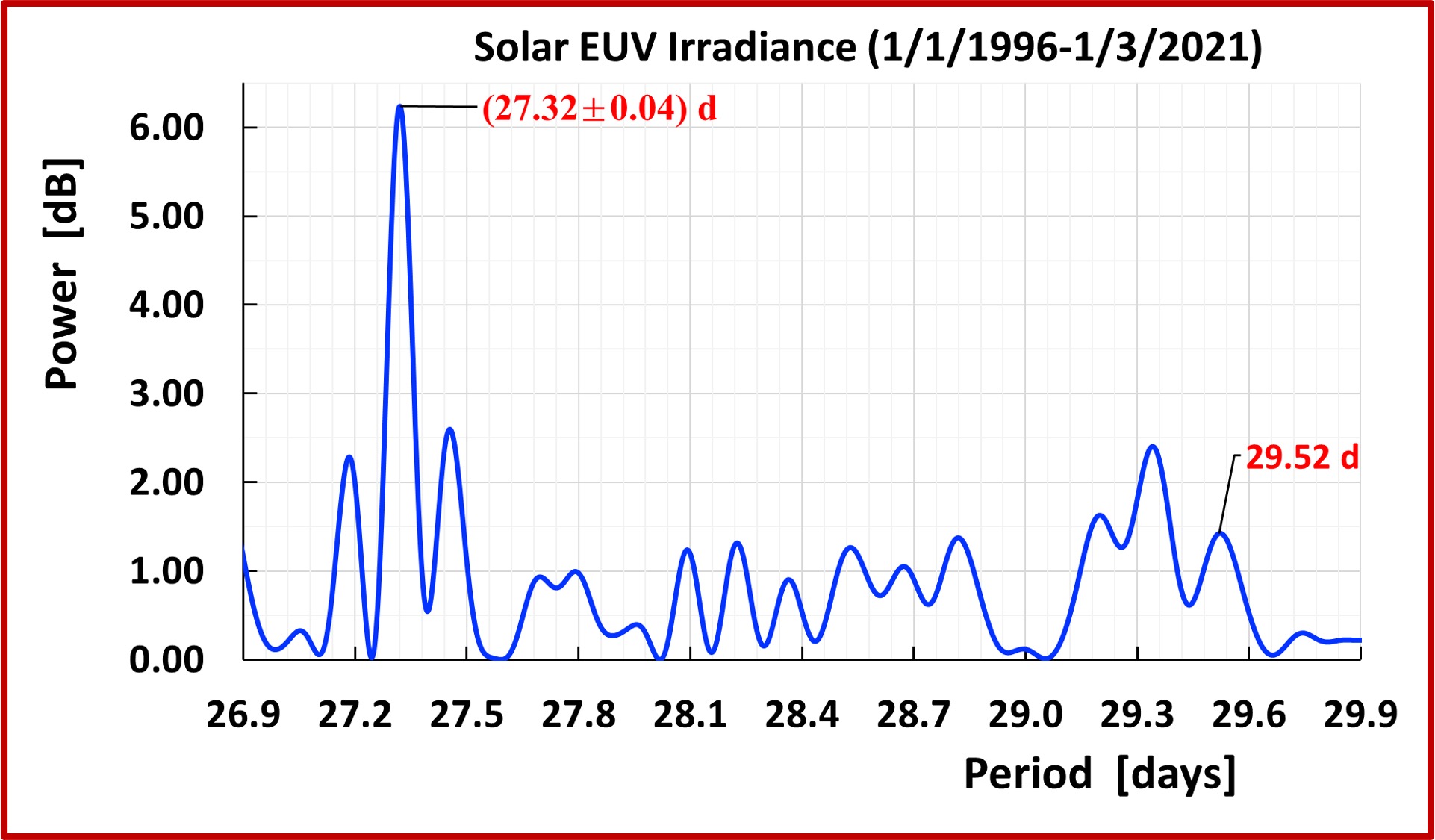}
    \caption{Fourier periodogram of daily solar EUV emission above 25 eV from 01/01/1996 to 01/03/2021, zoomed in on the 27–30 day range. The peak at 27.32 d (29.53 d) corresponds to the lunar sidereal (synodic) orbital rhythm. Contrary to expectation, the dominant driver of solar EUV emission appears to be of exo-solar origin (compare with Fig.~\ref{fig:tec} where the relative peak ratio 29.53:27.32 is different). See also \cite{Maroudas_phd, Appleton_proceedings, Lean_ionospheric, Rich_27day}.}
    \label{fig:euv}
\end{figure}

The EUV radiation is a direct manifestation of the solar corona, characterized by a steep temperature rise and density drop at an altitude of $\sim$2000 km above the photosphere. This sharp transition suggests that the solar atmosphere is being irradiated externally with a large effective cross-section. Fig.~\ref{fig:euv} presents the Fourier spectrum of solar EUV emissions, focusing on periodicities between 27 and 30 days. Remarkably, the sidereal lunar peak at $(27.32 \pm 0.04)$ days dominates, while the synodic lunar period (29.53 days) i.e. fixed to the Sun, is statistically insignificant.  

This dominant 27.32-day signal strongly suggests an exo-solar influence, possibly linked to planetary dynamics, particularly the Moon-Earth system. Axion antiquark nuggets (AQNs) \cite{Zhitnitsky_nonbaryonic, Zhitnitsky_solar_flares}, independently proposed as a viable DM candidate, provide the best available theoretical framework to support also this planetary dependency \cite{Zhitnitsky_mysterious}.  

It is important to note that solar EUV radiation is measured across the entire solar disk. Given the Sun's differential rotation (ranging from 25 to 30 days), there is no apparent reason to focus on the specific 27.32-day periodicity—except that it matches the lunar sidereal rhythm. In contrast, the Carrington rotation period is relevant only to low-latitude solar activity.  

Several solar phenomena remain unexplained within the standard solar model, and many exhibit planetary dependencies. These anomalies could serve as potential signatures of streaming DM, echoing Zwicky’s reasoning but applied locally within our solar system. Together, these independent anomalies strengthen the hypothesis of this work, highlighting the tension between observed solar behavior and theoretical expectations.

This conclusion gains further support when considering terrestrial anomalies, such as variations in the Earth’s atmosphere and seismic activity (see below). The wide range of unexplained local phenomena makes them strong candidates for DM signatures, as no other explanation can account for them simultaneously.

\subsection{Ionospheric Anomaly}

A long-standing and unresolved anomaly in atmospheric science is the "global annual anomaly," where the peak electron density in the ionosphere is approximately 25\% higher in December than in June. This effect, first noted in the 1930s \cite{Appleton_proceedings, Lean_ionospheric}, reflects an asymmetry between the December and June solstices.

Unexpectedly, the total electron content (TEC) of the global ionosphere also displays planetary dependencies, particularly with respect to Mercury and Venus, further complicating the mystery. Since the ionosphere is primarily influenced by solar UV-EUV irradiation, the persistence of this anomaly raises the question: Why has the solar corona not yet reached thermal equilibrium after $\sim$4.5 Gyr?

\begin{figure}[!htb]
    \centering
    \includegraphics[width=0.65\linewidth]{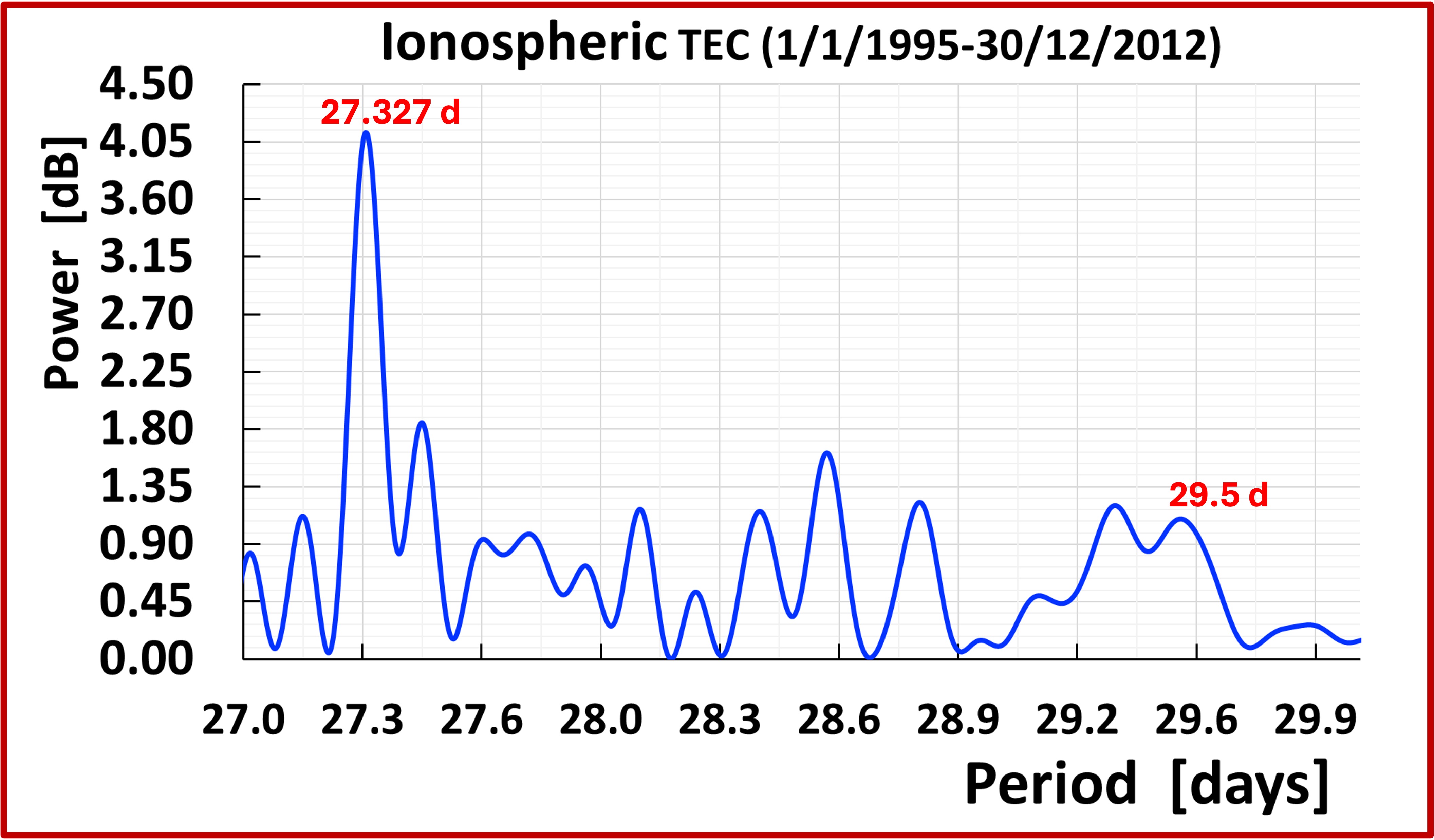}
    \caption{Fourier periodogram of daily ionospheric Total Electron Content (TEC) from 1/Jan/1995 to 30/Dec/2012, zoomed in on the 27–30 day range. The peak at 27.3 d (29.53 d) corresponds to the lunar sidereal (synodic) orbital rhythm. Contrary to expectation, the dominant driver of ionospheric TEC appears to be of exo-solar origin (compare with Fig.~\ref{fig:euv} where the relative peak ratio 29.53:27.32 is different). See \cite{Appleton_proceedings, Lean_ionospheric, Rich_27day}.}
    \label{fig:tec}
\end{figure}

The TEC values in December and June are $\sim2.87 \times 10^{32}$ and $\sim2.12 \times 10^{32}$ electrons, respectively. Fig.~\ref{fig:tec} presents the Fourier analysis of ionospheric TEC, highlighting a dominant sidereal lunar peak at 27.32 days, significantly stronger than the synodic peak at 29.53 days. The relative strengths of these peaks cannot be predicted a priori.  

Crucially, this suggests that a substantial portion of the ionospheric TEC is derived from an external source beyond the solar system. The most plausible explanation remains an incident stream of DM, particularly in the form of AQNs, a hypothesis proposed independently of this study (see also \cite{Zhitnitsky_mysterious}).

As noted in \cite{Rich_27day}, “Since solar EUV creates the ionized gas that composes the ionosphere, it seems obvious that a daily variation in solar EUV should result to daily variations in the topside plasma density and temperature. However, such a correlation is not found in the data!” This observation aligns perfectly with the current findings, reinforcing the idea that the dominant 27.32-day periodicity is not of solar origin. 

\subsection{Stratospheric temperature anomalies}

Stratospheric temperature anomalies have been observed annually around early January \cite{Zioutas_stratospheric}. Decades of data reveal an anomalous temperature increase in the upper stratosphere (38.5–47.5 km) in the northern hemisphere. Additionally, Fig.~\ref{fig:tec} shows a sidereal lunar dependence with a periodicity of approximately 27.32 days. This finding should be considered alongside the peak distribution in heliocentric longitudes (see Fig.~8 in Ref. \cite{Zioutas_stratospheric}).

Interestingly, this peak in the upper stratosphere disappears about 15 km below, at a height of 16 to 31 km. Of note, the relative ratio of the two peaks, 29.53d:27.32d, varies across neighboring regions of Earth's atmosphere, with the dominant peak at 27.32d suggesting an exo-solar origin. This likely points to streaming DM composed of different constituents, each interacting with varying strengths in the ionosphere and upper stratosphere. 

More strikingly, Fig.~8 in \cite{Zioutas_stratospheric} clearly shows that the peak in stratospheric temperature coincides with the planetary positions of Mercury and Venus as they propagate through heliocentric longitudes between \SI{90}{\degree} and \SI{270}{\degree}. The annual stratospheric anomaly strengthens, while the temperature peak disappears when the two inner planets propagate within longitudes between \SI{270}{\degree} and \SI{90}{\degree}. This unexpected planetary dependence on the two innermost planets reveals a double anomaly, reinforcing a planetary connection. This behavior mirrors Zwicky’s reasoning, but in this case, it occurs approximately 50 km above Earth's surface. The atmosphere has been continuously monitored for decades, making it a novel, low-threshold detector for the dark Universe. It offers built-in spatiotemporal resolution, with the Sun acting as a temporal signal amplifier (see Refs. \cite{Zioutas_stratospheric, Maroudas_phd}).

\begin{figure}[!htb]
    \centering
    \includegraphics[width=0.65\linewidth]{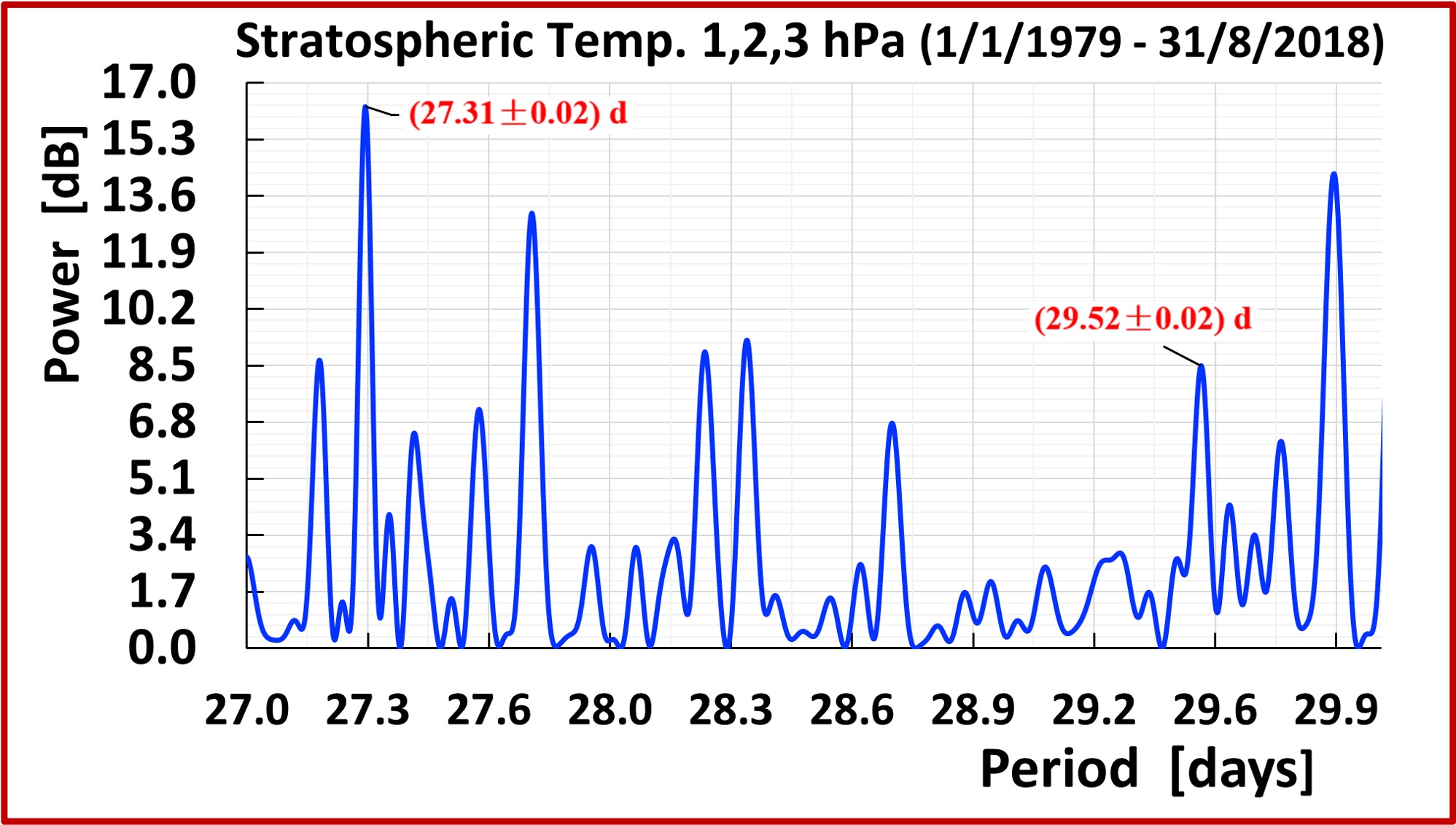}
    \caption{Fourier periodogram of the daily temperature in the upper stratosphere between 01/01/1979 - 31/08/2018, at a height of (38.5–47.5) km. This spectrum covers the region between 27–30 days. Here, the relative ratio of the peak heights 27.3 d/29.5 d appears to favor the 29.5 days synodic periodicity (=Moon month) compared to that for the TEC (see Figs.~\ref{fig:euv} and \ref{fig:tec} where the relative peak ratio 29.53:27.32 is different.) \cite{Maroudas_phd}.}
    \label{fig:strato}
\end{figure}

\subsection{Earthquakes}

\begin{figure}[!htb]
    \centering
    \includegraphics[width=0.65\linewidth]{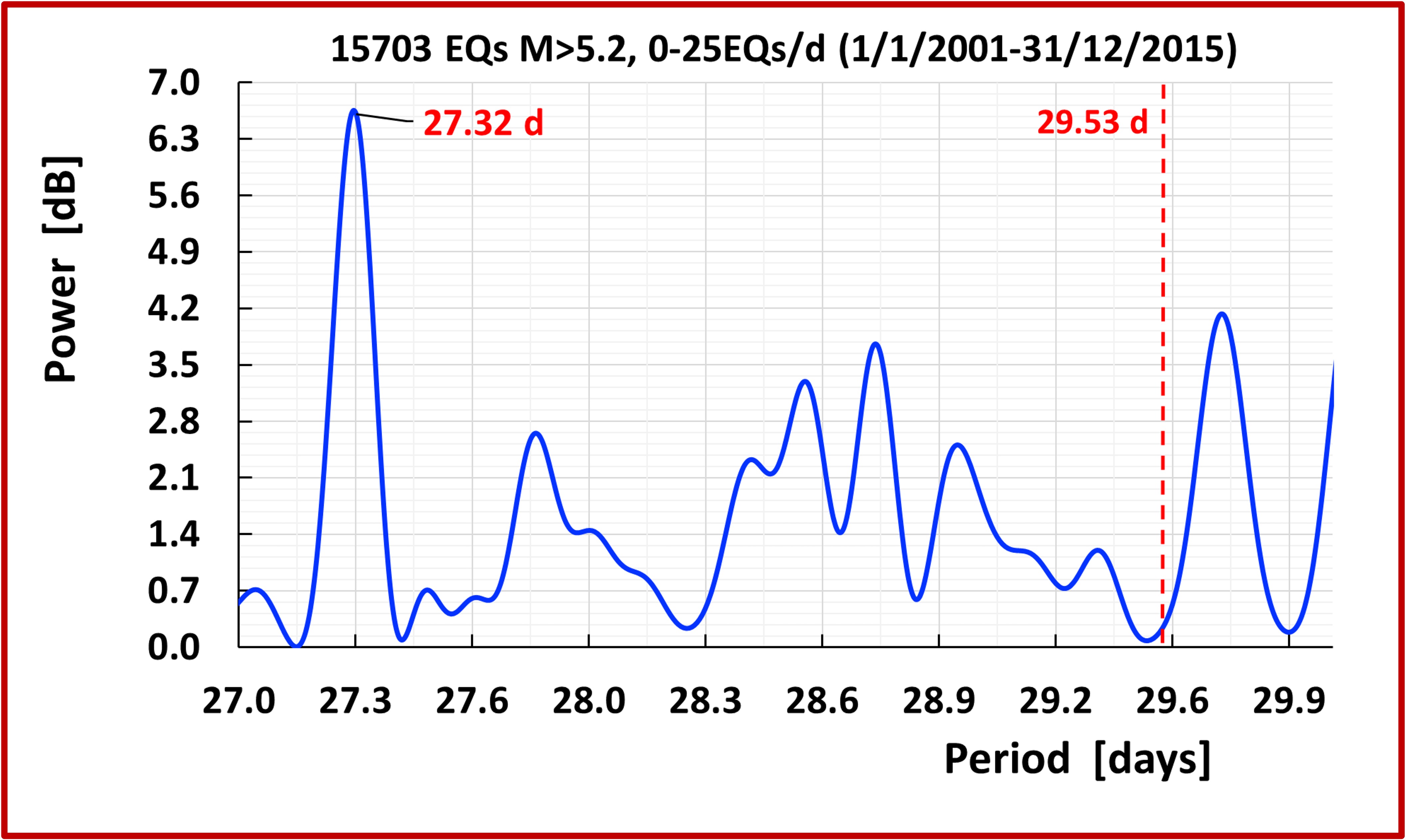}
    \caption{Fourier periodogram of the number of Earthquakes (EQs) of magnitude $M>5.2$, with a maximum of 25 EQs/day (01/01/2001 to 31/12/2015). Number of EQs = 15,703. The sidereal lunar rhythm is at (27.32 ± 0.05) days; the Moon month (synodic) at 29.53 days is absent \cite{Maroudas_phd}.}
    \label{fig:eq}
\end{figure}

Earthquakes (EQs) with magnitudes $M>5.2$ exhibit a sidereal lunar periodicity when selecting the occurrence of less than 25 EQs on a given day. The Fourier analysis in Fig.~\ref{fig:eq} reveals the sidereal lunar rhythm around the Earth, but not the synodic lunar cycle (29.53 days). Remarkably, this suggests that the occurrence of deep underground EQs is related to an external source beyond our solar system. In this context, Ref. \cite{Budker_infrasonic} explores the potential involvement of such a source in triggering EQs.

Fig.~\ref{fig:eq} presents the Fourier spectrum zoomed in on the range between 27 and 30 days. The analysis shows a dominant peak at 27.32 ± 0.05 days, corresponding to the sidereal lunar periodicity, while the synodic lunar period (29.53 days) is absent. The relative strength of these peaks cannot be predicted due to the unknown properties, such as interaction cross-sections and flux, of the assumed DM particles.

The correlation between Total Electron Content (TEC) and major EQs with magnitudes $M>8$ is noteworthy (see Fig. 7 in Ref. \cite{Zioutas_atmospheric}), revealing a clear relationship between EQs and global atmospheric plasma. This correlation extends to the global ionosphere, showing a 2D connection to the location of the catastrophic EQ. As a societal byproduct, this could provide a new early warning system, offering a time window of approximately two months before a large-magnitude EQ occurs.

We note that the EQs exhibit a stronger sidereal lunar periodicity at 27.32 days and very little at 29.53 days, i.e., the Moon month (synodic periodicity). It is worth recalling that the sidereal periodicity implies an exo-solar cause. In this context, we consider the possible influence of AQNs (anti-quark nuggets), which is particularly intriguing as it affects their impact on the atmosphere and the dynamical inner Earth (see also Ref. \cite{Budker_infrasonic}). AQNs most likely cannot provide the spatiotemporal energy release associated with an EQ, but they could serve as a trigger provoking an EQ in a mechanically stressed inner Earth. Thus, with EQs, we are likely dealing with events whose origin lies beyond the solar system, further evidenced by their planetary dependencies. The planetary influence on EQs has been observed when comparing their timestamps with planetary orbital positions \cite{Maroudas_phd}. Throughout this work, the Fourier spectrum serves as a common signature for exo-solar triggers, applied to various observables.

\subsection{Other dark matter signatures}

\begin{itemize}
    \item \textbf{Solar elemental abundance} has long been characterized as anomalous within standard solar physics (see Ref. \cite{Zioutas_invisible}). When the timing of this observable is projected onto heliocentric longitudes, the resulting planetary dependence further intensifies the observed anomaly, adding to its mystery. A statement from the cover page of the Journal *New Scientist* is worth mentioning \cite{Palus2017, sciencealert_2023}: "Something strange is going on inside the Sun." In fact, it has long been observed that the ratio of the coronal elemental composition is higher than the photospheric one, an effect that can not be explained by experimental uncertainties \cite{Perryman_gaia}. Additionally, a planetary dependence appears when projecting the First Ionization Potential (FIP) bias value on planetary orbital positions like that of Earth, or by combining Mercury and Venus \cite{Maroudas_phd}. The observed amplitudes are quite high, between 14\% and 18\% defying conventional explanations.
    
    \item \textbf{Magnetic bright points} (MBPs) play an important role in solar physics as a type of very small sunspots. They also exhibit planetary dependence when projecting their solar surface density on heliocentric longitudes of Venus and Earth (see Ref. \cite{Zioutas_invisible}), with the amplitude being similarly large (10–14\%).
    
    \item The diverse planetary dependencies observed with solar and terrestrial observables are also suggestive for \textbf{biomedicine}, where otherwise inexplicable phenomena like cancer have been reported \cite{Zioutas_melanoma, Zioutas_response, Zioutas_27_melanoma}, along with other in vivo observations \cite{Zioutas_births}. The dominant sidereal rhythm at 27.32 days, as well as single or synodic planetary dependencies (e.g., the 237 days Jupiter–Venus synod and the triple synod Jupiter–Earth–Venus), are associated with in vivo measurements, suggesting that the cause of a wide range of conditions in biomedicine might be of exo-solar origin. The widely assumed cause for melanoma is exposure to solar irradiation in the UV-EUV range, but its importance can only be speculated at present. This is particularly true since diagnoses tend to flare up during local summertime in both hemispheres \cite{Zioutas_melanoma, Zioutas_response, Zioutas_27_melanoma}. Interestingly, the diagnosis rate clearly increases in raw data during local summertime. It is the observed planetary dependencies that point to an exo-solar origin. Furthermore, observed multiple planetary dependencies confirm that normal, non-malignant behavior in vivo \cite{Zioutas_births} is influenced by an external factor beyond our solar system.
\end{itemize}

The only viable explanation for all these observations is streaming DM, which can interact strongly with living matter. Recent investigations with living matter have shown that it senses external impact with modulations like 27.32 days (sidereal lunar rhythm), 237 days (Jupiter–Venus synod), and 11 years (Jupiter–Earth–Venus synod) \cite{Zioutas_births, Zioutas_11years}. Each observation is statistically significant above $5\sigma$, and when combined, they reinforce the inital hypothesis that the unexpected planetary dependencies observed in vivo are real. This makes the anomalous behavior in biomedicine even more enigmatic. The open question remains whether such planetary dependencies also manifest in vitro.

\section{Discussion and Conclusion}

The central question of this work is whether the persistent anomalies and mysteries within the solar system may be the unnoticed manifestation of the dark Universe. Several key observations support the streaming DM scenario, extending Zwicky’s reasoning on a cosmic scale to local observations that are anomalous or mysterious even within the framework of known physics.

At smaller scales, tensions persist within current physical models, making it increasingly difficult to disregard the potential involvement of streaming DM in our vicinity. The widely accepted speed of DM constituents is around $~0.001c$ (where $c$ is the speed of light in vacuum), meaning that solar system bodies could serve as effective gravitational lenses for DM streams or clusters. As previously emphasized, the gravitational influence of a celestial body follows the relation \cite{Hoffmann_Gravitational, Patla_flux}: $1/(\text{DM velocity})^2$.

The inconsistencies addressed in this work may represent significant evidence that the current DM paradigm requires revision to explain these persistent anomalies. While cosmological models often overlook such effects, the gravitational influence exerted by orbiting solar system objects on incoming DM constituents could be critical for understanding phenomena that remain unexplained within the solar system.

The proper observations and data may help identify potential candidates from the dark sector. In particular, some of the solar system's anomalies and mysteries could be manifestations of the dark Universe. The unexpected planetary dependencies observed with solar system observables could offer an alternative to Zwicky's overestimated visible mass at the cosmic scale. The planetary dependence aligns with Zwicky's original tension regarding visible mass, reinforcing the need for a revised understanding of the underlying forces.

The long-term serial observations of the solar system, as presented in this work, mirror Zwicky's reasoning. All of the previously anomalous observations also exhibit one or more unexpected planetary dependencies. Zwicky’s discovery of "dunkle Materie" (dark matter) stemmed from the tension created by the overestimation of visible mass at the cosmic scale. In this work, the analogous tension is the overlooked planetary dependency, alongside the mysterious nature of the phenomena in question. These dependencies remain unexplained within the boundaries of known physics. As Frank Wilczek famously stated during a seminar at CERN (Nobel 2004): “Focus on anomalies and mysteries.” By aligning these solar mysteries with planetary orbital positions, this work adheres to Wilczek’s recommendation and invites further exploration.

\section*{Acknowledgments}
Y. K. S acknowledges support by IBS-R017-D1 of the Republic of Korea. This article is based on the work from COST Action COSMIC WISPers CA21106, supported by COST (European Cooperation in Science and Technology).

\section*{Conflict of interest}
The authors declare no conflict of interest.

\section*{Data availability}
The data supporting the findings of this study are available from the corresponding author upon reasonable request. 

\bibliography{refs}

\end{document}